# All-optical Hall effect by the dynamic toroidal moment in a cavity-based metamaterial

Zheng-Gao Dong,[1,2,3,*] Jie Zhu,[2] Xiaobo Yin,[2] Jiaqi Li,[1,3] Changgui Lu,[2] and Xiang Zhang[2]

[1]*Physics Department, Southeast University, Nanjing 211189, China*

[2]*5130 Etcheverry Hall, Nanoscale Science and Engineering Center, University of California, Berkeley, California 94720-1740, USA*

[3]*Research Center of Converging Technology, Southeast University, Nanjing 210096, People's Republic of China*

Dynamic dipolar toroidal response is demonstrated by an optical plasmonic metamaterial composed of double disks. This response with a hotspot of localized *E*-field concentration is a well-behaved toroidal cavity mode that exhibits a large Purcell factor due to its deep-subwavelength mode volume. All-optical Hall effect (photovoltaic) due to this optical toroidal moment is demonstrated numerically, in mimicking the magnetoelectric effect in multiferroic systems. The result shows a promising avenue to explore various optical phenomena associated with this intriguing dynamic toroidal moment.

PACS: 42.70.Qs, 41.20.Jb, 78.20.Ci, 73.20.Mf

A standard multipole expansion method conventionally decomposes the scattering field into contributions of electric and magnetic dipoles, quadrupoles, and so forth [1]. As is known, electric polarization breaks the space-inversion symmetry while magnetization breaks the time-reversal symmetry, and thus they are incomplete when confronting systems in particle physics with simultaneous violations of space-inversion and time-reversal symmetries. In 1957, Dipolar toroidal moment $\vec{T}$ was introduced to interpret the parity violation in weak interactions [2], which is now widely acknowledged not only in nucleons [3], atoms [4], molecules [5], and other elementary particles [6], but even in condensed matters such as multiferroic materials [7-9]. $\vec{T}$ is sometimes called an anapole moment, though they are different in a strict sense because the former can radiate while the latter is nonradiating (in the stationary limit they are exactly the same thing). As a classical configuration with $\vec{T}$, we can consider a conducting solenoid bent into a closed toroid with a poloidal current flowing in [2,9,10,11]. In condensed matters, a static $\vec{T}$ characterized by a vortex distribution of head-to-tail magnetic dipoles is associated with various intriguing properties, such as the magnetoelectric effect [9,10], dichroism [12,13], and nonreciprocal refraction [14].

However, these theoretically predicted properties associated with the elusive $\vec{T}$ are very weak to be observed experimentally from naturally occurring permanent multiferroic materials. Usually, they are readily masked by conventional multipole responses. Recently, dynamic $\vec{T}$ realized in metamaterials [12,15-22] may provide a platform to unveil the intriguing optics associated with it. As evidenced by the past decade, various optical phenomena have been demonstrated by metamaterials such as the negative refraction [23], super-imaging [24], plasmonically induced transparency [25], and cloaking [26]. From metamaterials with $\vec{T}$, interesting characteristics have been verified at microwaves as well as the infrared regime [12,15-22]. However, these sophisticate toroidal structures have certain geometrical asymmetries, and thus are hard for nanofabrication.

In this paper, before going to address the issue of designing an improved toroidal metamaterial that exhibits unique optical phenomena; it is of benefit to briefly review two elementary metamaterials. One is the cut-wire structure, scattering the field in the way of an electric dipole if the incident light is linearly polarized in the length direction of the wire [27]. The other is the split-ring structure, exhibiting a magnetic-dipole-like response when the light is incident with the *H*-field polarization parallel to the normal of the split ring [28]. Since electric and magnetic dipoles are excited by parallel *E*- and *H*-field components respectively, an optical $\vec{T}$, characterized by a centrally confined *E* field perpendicular to and attributed to the *H*-vortex distribution, could be induced by an incidence polarized parallel to $\vec{T}$. To mimic the classical toroidal configuration (namely, a current-carrying toroidal solenoid) [2,9,10,11], we demonstrate that a metallic metamaterial comprising strongly-coupled double disks (DD) is a qualified analogy of natural systems with $\vec{T}$.

Figure 1(a) schematically shows the excitation configuration of the proposed DD metamaterial (approximately a gap-bearing toroid)). The incident light propagates along the *x*-direction ($\vec{k}_{\text{light}} = k_x \hat{x}$) and is polarized perpendicular to the disk ($\vec{E}_{light} = E_z \hat{z}$) in order to satisfy $\vec{E}_{light} // \vec{T}$, which can excite a toroidal mode with the induced current density ($\vec{J}$) shown in Fig. 1(b). To eliminate the influence of unnecessary factors, the disks are set in the vacuum unless it will be specified. For simulations performed by the CST Microwave Studio, silver disks are considered with a dispersive permittivity $\varepsilon(\omega) = \varepsilon_\infty - \omega_p^2/(\omega^2 + i\omega\gamma)$, where the high-frequency permittivity $\varepsilon(\infty) = 6.0$, the plasma frequency $\omega_p = 1.5 \times 10^{16}$ rad/s, and the collision frequency $\gamma = 7.73 \times 10^{13}$ rad/s [29,30].

Spectra shown in Fig. 2 are calculated based on a two-dimensional DD array that is single layered in the *x*-direction but with periodicities of 1500 nm and 490 nm in *y*- and *z*- directions, respectively. According to the *H*-field map presented in insets of Fig. 2, the first mode at 63.8 THz is a dipolar magnetic resonance, which was extensively studied in magnetic metamaterials comprising the split-ring or double-bar structure [31,32]. The second mode around 105 THz is a quadrupole magnetic resonance [33]. However, the third mode at 130 THz, characterized by the *H*-vortex as well as the $\vec{J}$ distribution shown in Fig. 1(b), was rarely (if not never) discussed in literatures. To the interest of this work, it will be demonstrated that this response has a dominant $\vec{T}$ and thus is highlighted for explorations of intriguing optical properties. Note that reflectance in Fig. 2 is negligible for the toroidal resonance as well as for dipolar and quadrupole magnetic modes. This implies the weak scattering characteristic of $\vec{T}$ [12, 34], for which any decrease in the transmission is primarily caused by the absorption instead of the scattering. Such a nonscattering characteristic is in contrast to that of previously designed toroidal metamaterials, where reflection peaks in spectra manifested themselves as strongly scattering responses. To obtain the toroidal response with an obvious *H*-vortex distribution, the DD gap must be small enough to ensure that paired disks are in strong plasmon coupling. Additionally, the scattering/absorption ratio increases with the disk thickness as well as the gap, while the toroidal mode itself suffers a redshift with the reduction of the disk thickness (the toroidal resonant frequency is independent of the disk thickness for disks thicker than 40 nm due to the skin effect).

In principle, the transmission/reflection spectra can be decomposed into radiating contributions of multipoles, including the electric, magnetic, and toroidal ones. The $\vec{J}$ distribution in disks is used to analyze these resonant responses in terms of multipole scattering contributions with [15,17,35] dipolar electric moment: $\vec{P} = \dfrac{1}{i\omega}\int \vec{J}d^3r$, dipolar magnetic moment: $\vec{M} = \dfrac{1}{2c}\int (\vec{r}\times\vec{J})d^3r$, dipolar toroidal moment: $\vec{T} = \dfrac{1}{10c}\int [(\vec{r}\cdot\vec{J})\vec{r} - 2r^2\vec{J}]d^3r$, quadrupole electric moment: $Q_{\alpha\beta} = \dfrac{1}{i2\omega}\int [r_\alpha J_\beta + r_\beta J_\alpha - \dfrac{2}{3}(\vec{r}\cdot\vec{J})\delta_{\alpha\beta}]d^3r$,

quadrupole magnetic moment: $M_{\alpha\beta} = \dfrac{1}{3c}\int [(\vec{r}\times\vec{J})_\alpha r_\beta + (\vec{r}\times\vec{J})_\beta r_\alpha]d^3r$, where $c$ is the speed of light, $\vec{r}$ is the distance vector from the origin (DD gap center) to point $(x, y, z)$ in a Cartesian coordinate, and $\alpha, \beta = x, y, z$. Accordingly, the decomposed far-field scattering power by these multipole moments can be written as $I_P = 2\omega^4|\vec{P}|^2/3c^3$, $I_M = 2\omega^4|\vec{M}|^2/3c^3$, $I_T = 2\omega^6|\vec{T}|^2/3c^5$, $I_Q^e = \dfrac{\omega^6}{5c^5}\sum|Q_{\alpha\beta}|^2$, and $I_Q^m = \dfrac{\omega^6}{40c^5}\sum|M_{\alpha\beta}|^2$. For the resonance at 130 THz, a dominant scattering power

attributed to the toroidal moment $T_z$ is found from Fig. 3, which cannot be accounted by common electric and magnetic multipole moments, and thus confirms the unambiguous toroidal mode. The accompanied weak scattering peak from the electric dipole $P_z$ should be a result of the *gap-bearing* feature of toroid-like DD. Additionally, the toroidal mode in Fig. 3 has a wide profile extending beyond 140 THz, not matching the narrow absorption peak around 130 THz in Fig. 2. This is due to a nonresonant current distribution similar to the toroidal current shown in Fig. 1(b), extending in a wide frequency range.

It is interesting to investigate the performance of this toroidal resonance as a cavity mode, since the field confinement in this toroidal cavity (Fig. 4) is different from common cavity modes, such as a whispering-gallery or a Fabry-Perot cavity. This toroidal cavity mode can not only concentrate the *E* field into a hotspot at the DD center [Fig. 4(a)], but also confine the *H* field in a vortex distribution [Fig. 4(b)]. As for the counterintuitive observation of field maps exhibiting a three-fold rotational symmetry, it is caused by the imperfectly formed *H*-vortex in the gap when the incident light comes along the *x*-direction, since the Ag/air/Ag structure is not in a deep subwavelength scale. Nevertheless, if an infinitesimal line-current source, oscillating in the *z*-direction, is set in the gap center of the DD, this counterintuitive observation will disappear and vanishing field can be radiated to the outside. According to the simulation, at the hotspot of the DD cavity, the probed maximum of the *E*-field magnitude is $|\vec{E}_{local}| = 28.5$ V/m [Fig. 4(a)], indicating an intensity enhancement more than 800 times since the light is incident with an *E*-field magnitude $|\vec{E}_{light}|=1.0$ V/m. Accordingly, the mode volume is calculated to be as small as $0.0011(\frac{\lambda_0}{2})^3$ by using the formula $V_{eff}=\int\varepsilon(\vec{r})|\vec{E}(\vec{r})|^2 d^3r / \varepsilon(\vec{r}_{max})\max(|\vec{E}(\vec{r})|^2)$, where $\lambda_0 = c/f_0 = 7.692$ μm is the freespace wavelength corresponding to the resonant frequency $f_0 = 130$ THz, $\vec{r}_{max}$ is the hotspot location, and $\varepsilon(\vec{r})=1$ for the vacuum gap. Considering that the quality factor ($Q$) is about 19.8 (determined by the resonant frequency $f_0$ divided by the full width at half maximum of the spectrum in Fig. 2), a Purcell factor $F_P = (3/4\pi^2)\lambda_0^3(Q/V_{eff})$ as large as $1.12\times10^4$ can be reached. Additionally, such a cavity can be pushed into a deep subwavelength scale if the gap is squeezed in further. For example, when the gap size is decreased to 4 nm, the frequency of the toroidal response will shift to 89 THz. The results verify that the toroidal cavity mode excited in the DD structure is promising to be used in situations of optical cavity. Also, the giant localized field concentration at the hotspot is attractive for enhanced light emissions and nonlinear effects.

In multiferroic systems, it is known that the $\vec{T}$, for which both the spatial and temporal inversion symmetries are broken, can lead to the magnetoelectric effect [9-11]. That is, an applied *H* field induces the appearance of a built-in *E* field in the orthogonal direction, and vice versa. As schematically presented in Fig. 5(a), a steady poloidal current homogeneously flowing on the toroid surface will lead to a uniformly distributed *H*-vortex (counter-clockwise), which is described by a $\vec{T}$ perpendicular to the vortex. If an external $\vec{B}$ field is applied along the *x*-direction [Fig. 5(b)], the poloidal current contours with magnetic dipoles unparallel to the applied $\vec{B}$ will be dragged transversally (i.e., shift to the left side of the toroid) in order to orient their magnetic dipoles in the direction of $\vec{B}$. Such a redistribution of the originally uniform carriers implies an unbalanced accumulation of carriers on the left side of the toroid, resulting in an electric polarization ($\vec{P}$) in the $x$-direction [9-11]. This phenomenon was observed experimentally in naturally occurring ferrotoroic materials, so-called the magnetoelectric effect ($\vec{P} \propto \vec{T} \times \vec{B}$). For the optical toroidal response demonstrated in this work, its moment is dynamic, instead of the stationary one in ferrotoroic materials. Nevertheless, a similar magnetoelectric effect can be obtained as well, regardless of its optically dynamic nature. The difference is that a DC polarization, fed from the incident light itself, can be induced without an externally applied $\vec{B}$ field.

For the Ag/air/Ag structure, no obvious magnetoelectric effect can be observed from the centrally-positioned $H$-vortex distribution [Fig. 5(c)]. However, it will emerge if a semiconductor with a high refractive index (e.g., silicon) is filled in the gap, characterized by the nonuniform and left-shifted $H$-vortex in Fig. 5(d). A direct result from the gap-filled Ag/Si/Ag structure is the resonant frequency lowering to 39 THz (approximately equal to $f_0/\sqrt{\varepsilon_{silicon}}$), which makes the sandwich structure in a deep subwavelength geometry (the disk diameter is now about 6.5 times smaller than the freespace wavelength of the toroidal resonance). Subsequently, the $H$-field component of the incident light can be dealt with quasistatic approximation, acting like the applied uniform magnetic field for a stationary magnetoelectric effect in multiferroics [9-11]. Moreover, since $\vec{T}$ is parallel to the incident $\vec{E}_{light}$ as specified earlier, it is interesting to find that the magnetoelectric relationship $\vec{P} \propto \vec{T} \times \vec{B}$ now means

$$\pm \vec{P} \mathbin{/\mkern-5mu/} (\vec{E}_{\text{light}} \times \vec{H}_{\text{light}}) \mathbin{/\mkern-5mu/} \vec{k}_{\text{light}},$$

where the sign depends on the positively or negatively charged carrier. Considering that electrons are the carrier in a metal, a plenty of unbalanced electrons will be accumulated on the left side of the disks although they are oscillating with $\vec{T}$. As a result, a DC polarization $\vec{P}$ along the light propagating direction ($\vec{P} \mathbin{/\mkern-5mu/} \vec{k}_{\text{light}}$) will be retrieved from the incident light. This intriguing phenomenon is verified by the simulation result in Fig. 5(d), where current contours dragged to the left side of the toroid are similar to that expected in static ferrotoroic systems [Fig. 5(b)]. On the other hand, it is apparent that this phenomenon can also be understood in the context of Hall effect, because accumulated electrons on the left side, due to Ampere forces on contours of circulating current, is basically a result of charge carriers flowing in a magnetic field. Although the photoinduced $V_H$ is a DC signal, it is pulsating with the phase of the incident wave (Fig. 6), thus acts as an all-optical full-wave rectifier (For optical frequencies, usually only the time-averaged value is considered). For this unprecedented all-optical Hall effect, it is different from the magnetoelectric effect in permanent multiferroic materials since it is self-induced by the incident light without the externally applied magnetic field. Meanwhile, it should also be distinguished from phenomena termed "optical spin Hall effect", "Hall effect of light", and "optical Hall effect" [36]. Usually, an optical Hall effect means a transversal shift of the light trajectory at an interface of two materials with different refractive indices.

Additionally, the magnetoelectric effect for ferrotoroic materials typically works both ways, which means an external electric field can also induce a magnetization. Therefore, it deserves a further investigation about this reversible effect in such a dynamic toroidal response.

To summarize, we proposed a plasmonic metamaterial with a dominant $\vec{T}$, attributed to the resonantly induced poloidal current. Considering that this toroidal mode is nonscattering as well as in a deep subwavelength scale, it can be regarded as an unprecedented optical anapole moment. The resultant toroidal cavity characteristic with a large Purcell factor, due to the $E$ field concentrated at the hotspot, is of great advantage as compared with common cavity modes. Particularly, similar to the magnetoelectric effect in ferrotoroic materials, all-optical Hall effect attributed to the dynamic $\vec{T}$ is demonstrated from the deep-subwavelength Ag/Si/Ag structure. In view of the enhancement characteristics of localized plasmon resonance and the orbital nature of the induced current (instead of a spin one), a plenty of enhanced optical phenomena connected with the elusive $\vec{T}$ deserve further explorations.

This work was supported by the National Natural Science Foundation of China (Nos. 11174051, 10904012, 11004026, and 61007018) and the US National Science Foundation (NSF) Nanoscale Science and Engineering Center CMMI-0751621.

*Electronic address: zgdong@seu.edu.cn

FIG. 1 (color online). (a) Primary metamaterial element with a response of dynamic anapole moment, where the radius $r = 600$ nm, thickness $t = 40$ nm, and gap $g = 10$ nm for the silver disk. The coordinate origin is positioned at the gap center. (b) The current distribution for the dipolar toroidal response in the DD (note that the gap between the DD is enlarged intentionally for eye guidance).

FIG. 2 (color online). Spectra of transmittance ($T$), reflectance ($R$), and absorptance ($A = 1 - T - R$) for the proposed metamaterial configuration. The inset presents maps of the $H$-field vector on the middle of the gap, corresponding to the three resonant modes in the frequency sequence. The vanishing reflectance implies well-behaved nonscattering characteristics of these modes (Note the reflectance curve is multiplied by a factor of 10, so that the peaks and their shapes are clear).

FIG. 3 (color online). Scattering powers of multipole moment. Note the dominant toroidal moment $T_z$ is responsible for the resonance around 130 THz.

FIG. 4 (color online). The electromagnetic concentration capability of the toroidal cavity mode in the DD structure. There are simultaneous enhancements for the $E$ field (a) as well as the $H$ field (b) in the DD gap.

FIG. 5 (color online). The amplitude distributions of tangential current density ($|\vec{J}_{//}| = \sqrt{J_x^2 + J_y^2}$) are plotted at 130 THz and 39 THz for the Ag/air/Ag and Ag/Si/Ag structures, respectively. A transversal displacement of the carriers (electrons) toward the left side can be verified (Dashed lines indicate the disk profiles).

FIG. 6 (color online). The photovoltaic signal by the all-optical Hall effect. Note the magnetic vortices on the upper panel are plotted only in a half cycle (at 39 THz), while the corresponding rectified Hall voltage is in a full cycle.

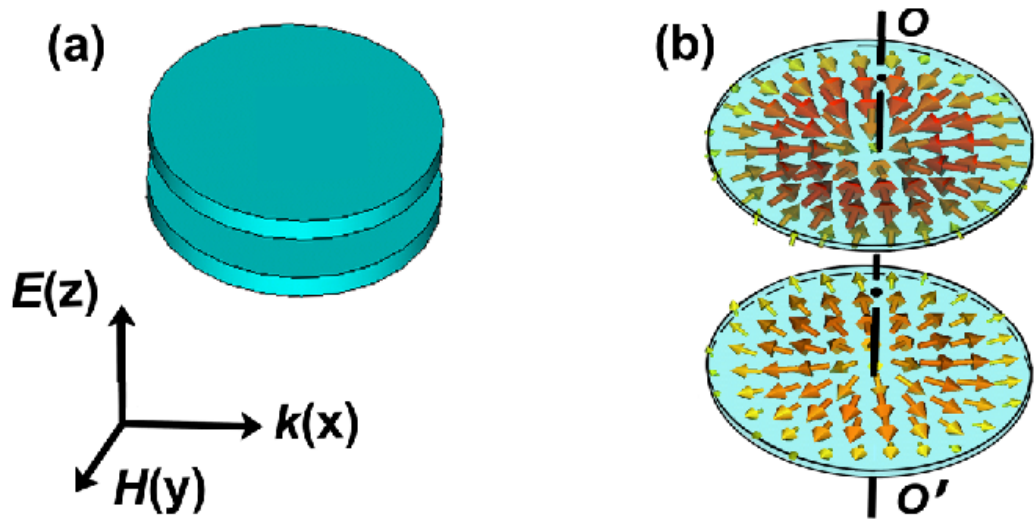

FIG. 1 Dong et al.

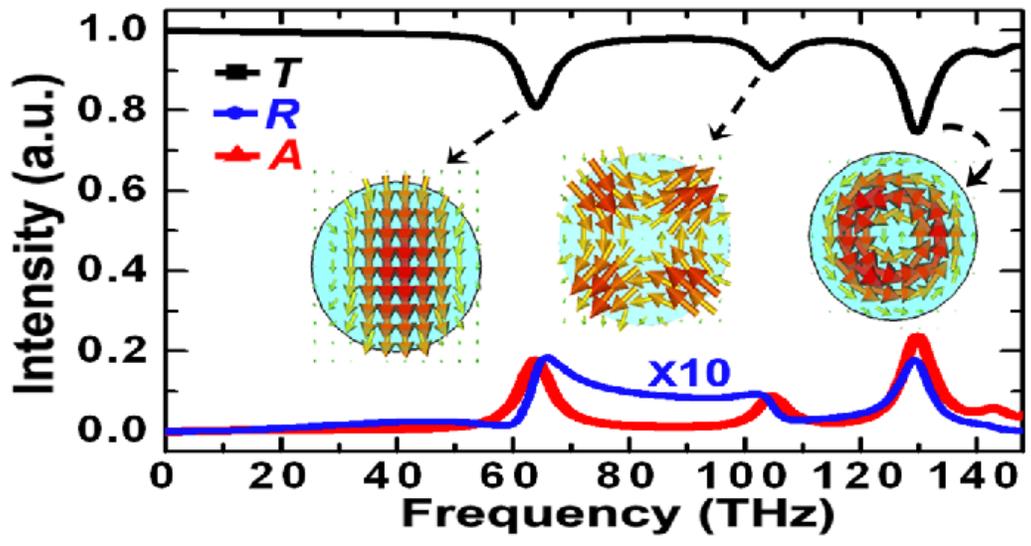

FIG. 2 Dong et al.

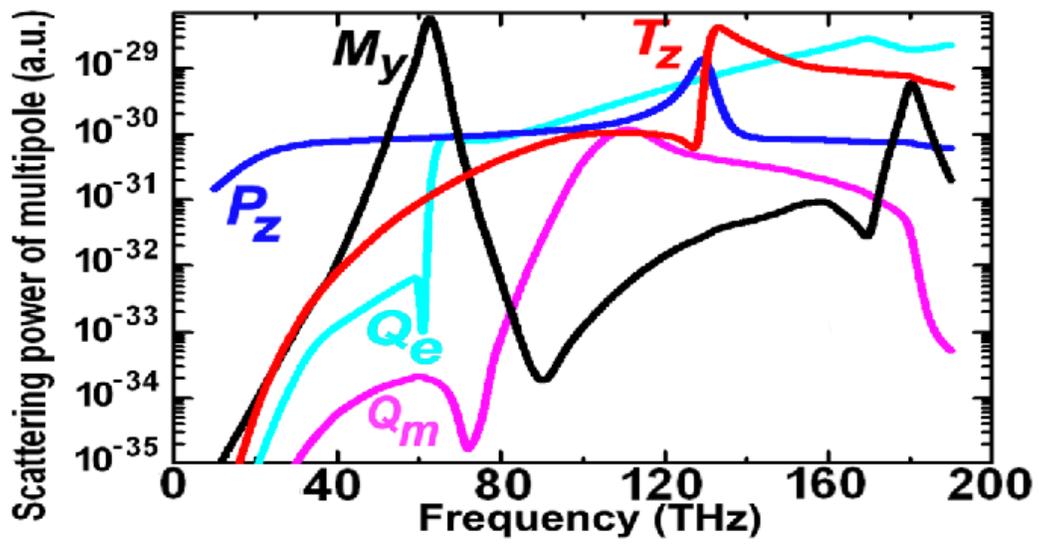

FIG. 3 Dong et al.

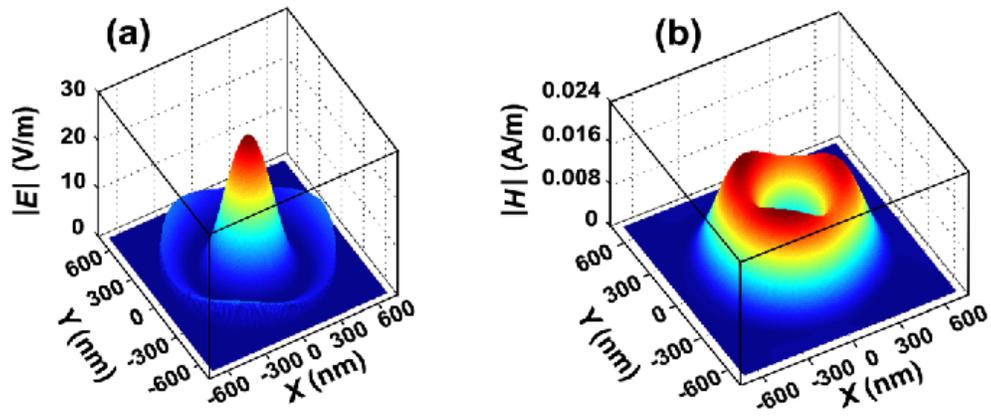

FIG. 4 Dong et al.

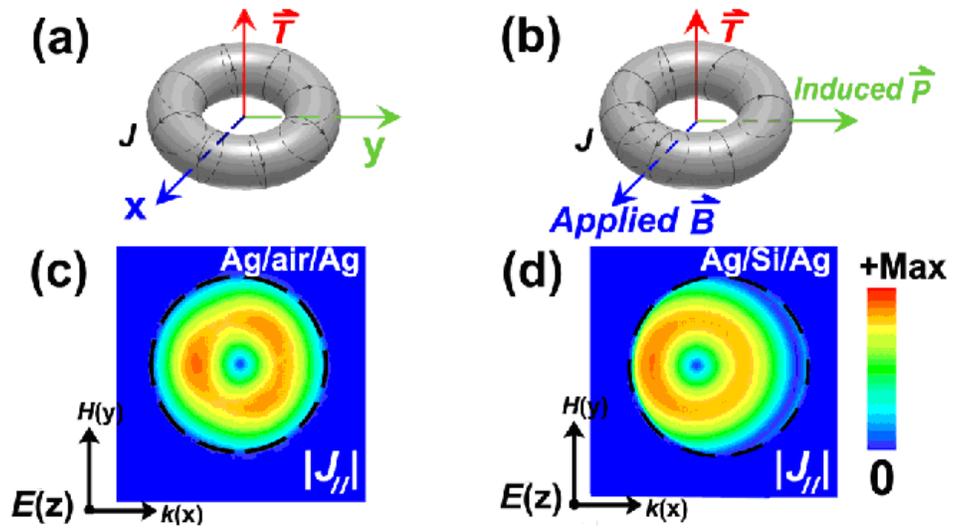

FIG. 5 Dong et al.

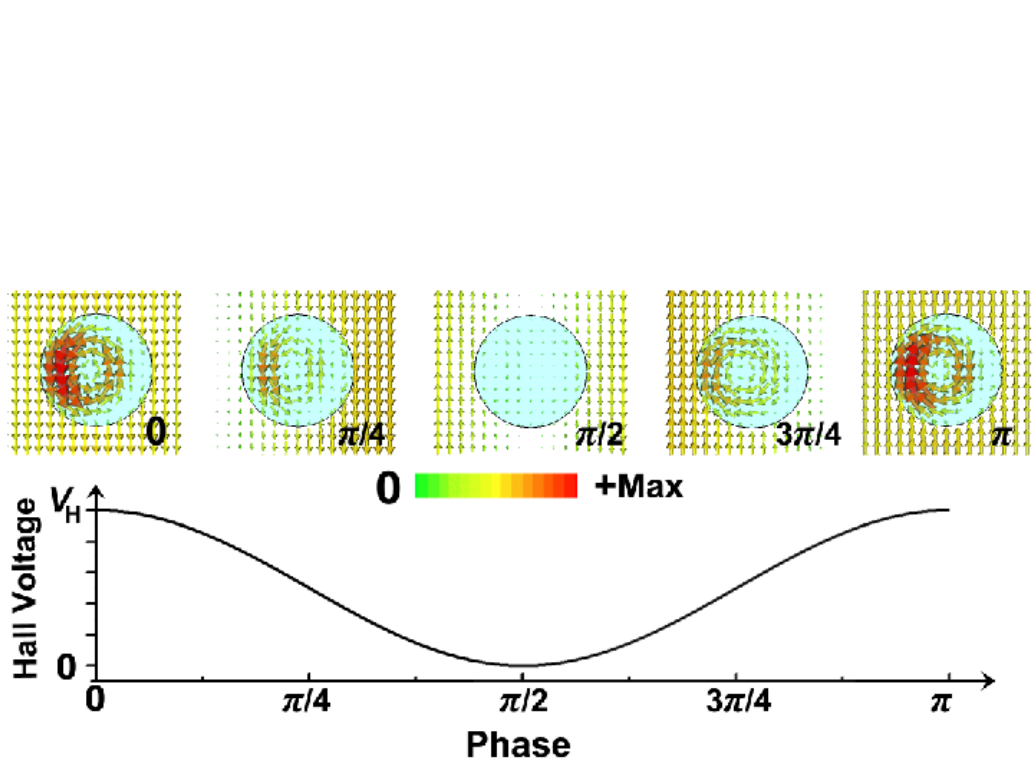

FIG. 6 Dong et al.